\documentclass{article}
\usepackage{amsmath,amssymb}
\usepackage{textcomp} 
\usepackage{rotating}
\usepackage{graphicx}
\usepackage{hyperref}
\usepackage{subfig}
\usepackage[right]{lineno}
\usepackage{setspace}
\usepackage{makecell}

\usepackage[aboveskip=1pt,labelfont=bf,labelsep=period,justification=raggedright,singlelinecheck=off]{caption}

\makeatletter
\renewcommand{\@biblabel}[1]{\quad#1.}
\makeatother

\date{}

\begin{document}

\begin{flushleft}
{\Large
\textbf\newline{Automatic acoustic identification of individual animals: Improving generalisation across species and recording conditions}
}
\newline
\\
Dan Stowell\textsuperscript{1*},
Tereza Petruskov{\'{a}}\textsuperscript{2},
Martin \v{S}\'{a}lek\textsuperscript{3,4},
Pavel Linhart\textsuperscript{5}
\\
\bigskip
\bf{1} Machine Listening Lab, Centre for Digital Music, Queen Mary University of London, UK
\\
\bf{2} Department of Ecology, Faculty of Science, Charles University, Prague, Czech Republic
\\
\bf{3} Institute of Vertebrate Biology, The Czech Academy of Sciences, Brno, Czech Republic
\\
\bf{4} Faculty of Environmental Sciences, Czech University of Life Sciences Prague, Prague, Czech Republic 
\\
\bf{5} Department of Behavioural Ecology, Faculty of Biology, Adam Mickiewicz University, Pozna\'{n}, Poland
\\
\bigskip

* dan.stowell@qmul.ac.uk

\end{flushleft}

\section*{Abstract}

Many animals emit vocal sounds which, independently from the sounds' function, embed some individually-distinctive signature. Thus the automatic recognition of individuals by sound is a potentially powerful tool for zoology and ecology research and practical monitoring. Here we present a general automatic identification method, that can work across multiple animal species with various levels of complexity in their communication systems. We further introduce new analysis techniques based on dataset manipulations that can evaluate the robustness and generality of a classifier. By using these techniques we confirmed the presence of experimental confounds in situations resembling those from past studies.
We introduce data manipulations that can reduce the impact of these confounds, compatible with \textit{any} classifier.
We suggest that assessment of confounds should become a standard part of future studies to ensure they do not report over-optimistic results. We provide annotated recordings used for analyses along with this study and we call for dataset sharing to be a common practice to enhance development of methods and comparisons of results.   

\vspace{2cm}

\textbf{Keywords:}
animal communication;
individual differences;
individuality;
acoustic monitoring;
song repertoire;
vocalisation.

\newpage

\newpage
\section{Introduction}

Animal vocalisations exhibit consistent individually-distinctive patterns, often referred to as acoustic signatures. Individual differences in acoustic signals have been reported universally across vertebrate species (e.g., fish \cite{amorim_variability_2008}, amphibians \cite{bee_neighbour-stranger_2001}, birds \cite{Terry:2005}, mammals \cite{taylor_contribution_2010}). Individual differences may arise from various sources, for example:
distinctive fundamental frequency and harmonic structure of acoustic signal can result from individual vocal tract anatomy \cite{taylor_contribution_2010,gamba_modeling_2017}; distinct temporal or frequency modulation patterns of vocal elements may result from inaccurate matching of innate or learned template or can occur \textit{de novo} through improvisation \cite{janik_vocal_1997}.
Such individual signatures provide individual recognition cues for other conspecific animals, and individual recognition based on acoustic signals is widespread among animals \cite{wiley_specificity_2013}.
Long-lasting individual recognition spanning over one or more year has also been often demonstrated \cite{boeckle_long-term_2012,insley_long-term_2000,Briefer:2012}.
External and internal factors such as, for example, sound degradation during transmission \cite{slabbekoorn_singing_2004,Mouterde:2014}, variable ambient temperature \cite{gambale_individual_2014}, inner motivation state \cite{collins_vocal_2004,linhart_being_2013}, acquisition of new sounds during life \cite{kroodsma_diversity_2004}, may potentially increase variation of acoustic signals. Despite these potential complications, robust individual signatures were found in many taxa.

Besides being studied for their crucial importance in social interactions \cite{thom_female_2012,bradbury_principles_1998,crowley_evolving_1996}, individual signatures can become a valuable tool for monitoring animals. Acoustic monitoring of individuals of various species based on vocal cues could become a powerful tool in conservation (reviewed in \cite{Terry:2005,mennill_individual_2011,blumstein_acoustic_2011}). Classical capture-mark methods of individual monitoring involve physically disturbing the animals of interest and might have a negative impact on health of studied animals or their behaviour (e.g. \cite{Johnsen:1997,Gervais:2006,Linhart:2012,rivera-gutierrez_songbirds_2015}). Also, concerns have been raised about possible biases in demographic and behavioural studies resulting from trap boldness or shyness of specific individuals \cite{camacho_lifelong_2017}. Individual acoustic monitoring offers a great advantage of being non-invasive, and thus can be deployed across species with fewer concerns about effect on behaviour \cite{Terry:2005}. It also may reveal complementary or more detailed information about species behaviour than classical methods \cite{Petruskova:2015,Laiolo:2007,kirschel_territorial_2011, Spillmann:2017}.

Despite many pilot studies \cite{delport_vocal_2002,Laiolo:2007,Adi:2010,terry_census_2002}, automatic acoustic individual identification is still not routinely applied. It is usually restricted to a particular research team or even to a single research project, and eventually, might be abandoned altogether for a particular species. Part of the problem probably lies in the fact that methods of acoustic individual identification were closely tailored to a single species (software platform, acoustic features used, etc.). This is good in order to obtain the best possible results for a particular species but it also hinders general, widespread application because methods need to be developed from scratch for each new species or even project. Little attention has been paid to developing general methods of automatic acoustic individual identification (henceforth ``AAII'') which could be used across different species.

A few studies in the past have proposed to develop a general, call-type-independent acoustic identification, working towards approaches that could be used across different species, having simple as well as complex vocalisations \cite{Fox:2008}. Despite promising results, most of the published papers included vocalisations recorded within very limited periods of time (a few hours in a day) \cite{Fox:2008,Fox:2008a,Cheng:2010,Cheng:2012}. Hence, these studies might have failed to separate effects of target signal and potentially confounding effects of particular recording conditions and background sound, which have been reported as notable problems in case of other machine learning tasks \cite{Szegedy:2013,Mesaros:2018dcase}. Reducing such confounds directly, by recording an animal in different backgrounds, may not be achievable in field conditions since animals typically live within limited home ranges and territories. However, acoustic background can change during the breeding season due to vegetation changes or cycles in activity of different bird species. Also, song birds may change territories in subsequent years or even within a single season \cite{Petruskova:2015}. 
Some other studies of individual acoustic identification, on the other hand, provided evidence that machine learning acoustic identification can be robust in respect to possible long-term changes in the acoustic background but did not provide evidence of being generally usable for multiple species \cite{Spillmann:2017,Adi:2010}. Therefore, the challenge of reliable generalisation of machine learning approach in acoustic individual identification across different conditions and different species has not yet been satisfactorily demonstrated.

\subsection{Previous methods for automatic classification of individuals from their vocalisations}
%
%
%
We briefly review studies representing methods for automatic classification of individuals.
Note that in the present work, as in many of the cited works, we set aside questions of automating the prior steps of recording focal birds and isolating the recording segments in which they are active.
It is common, in preparing data sets, for recordists to collate recordings and manually trim them to the regions containing the ``foreground'' individual of interest (often with some background noise), discarding the regions containing only background sound.
In the present work we will make use of both the foreground and background clips, and our method will be applicable whether such segmentation is done manually or automatically.

Matching a signal against a library of templates is a well-known bioacoustic technique, most commonly using spectrogram (sonogram) representations of the sound, via spectrogram cross-correlation \cite{Khanna:1997}.
For identifying individuals, template matching will work in principle when the individuals' vocalisations are strongly stereotyped with stable individual differences---and in practice this can give good recognition results for some species \cite{Foote:2013}.
However, template matching is only applicable to a minority of species. It is strongly call-type dependent and requires a library covering all of the vocalisation units that are to be identified. It is unlikely to be useful for species which have a very large vocabulary, high variability, or whose vocabulary changes substantially across seasons.

An approach which can be more independent of call type is that of Gaussian mixture models (GMMs), previously used extensively in human speech technology \cite{Ptacek:2016,Spillmann:2017}.
These do not rely on a strongly fixed template but rather build a statistical model summarising the observations (e.g.\ the spectral shapes) that are likely to be produced from each individual.
A particularly useful aspect of the GMM paradigm is that it can straightforwardly incorporate the concept of a ``universal background model'' (UBM),
which represents not ``background'' as ordinarily understood but a universal pool of the sounds that might be produced by individuals known and unknown.
It therefore allows for the practical possibility that a given sound might come from unknown individuals that are not part of the target set \cite{Ptacek:2016}.
This approach has been used in songbirds, although without testing across multiple seasons \cite{Ptacek:2016}, and for orangutan including across-season evaluation \cite{Spillmann:2017}.

The GMM is a very basic statistical model, which does not incorporate any notion of temporal structure. It thus misses out on making use of a large amount of information in the signal.
One way to improve on this, again well-developed in human speech technology, is to apply hidden Markov models (HMMs).
HMMs are statistical models of temporal structure and have more flexibility than template-matching.
However, in general they are likely to be call-type-dependent since they do encode the temporal structure observed in each vocalisation.
Adi et al.\ used HMMs for recognising individual songbirds, in this case ortolan buntings, with a pragmatic approach to call-type dependence \cite{Adi:2010}.
They first applied HMMs to infer the call type active in a given recording (independent of individual),
and then given the call type, applied GMMs to infer which individual was active.

Other computational approaches have been studied.
Cheng et al.\ compared four classifier methods, aiming to develop call-type-independent recognition across three passerine species \cite{Cheng:2012}.
They found HMM and support vector machines to be favourable among the methods they tested.
However, the data used in this study was relatively limited:
it was based on single recording sessions per individual, and thus could not test across-year performance;
and the authors deliberately curated the data to select clean recordings with minimal noise, acknowledging that this would not be representative of realistic recordings.
Fox et al.\ also focused on the challenge of call-independent identification, across three other passerine species \cite{Fox:2008a,Fox:2008}.
They used a neural network classifier, and achieved good performance for their species.
However, again the data for this study was based on a single session per individual,
which makes it unclear how far the findings generalise across days and years,
and also does not fully test whether the results may be affected by confounding factors such as recording conditions.

Computational methods for various automatic recognition tasks have recently been dominated and dramatically improved by new trends in machine learning, including deep learning. Within that broad field, the challenge of reliable generalisation is far from solved, and is an active research topic. Within bioacoustics this has recently been studied for detection of bird sounds \cite{Stowell:2018badchj}. In deep learning, it was discovered that even the best-performing deep neural networks might be surprisingly non-robust, and could be forced to change their decisions by the addition of tiny imperceptible amounts of background noise to an image \cite{Szegedy:2013}.

Note that deep learning systems also typically require very large amounts of data to train, meaning they may currently be infeasible for tasks such as acoustic individual ID in which the number of recordings per individual is necessarily limited.
For deep learning, ``data augmentation'' has been used to expand dataset sizes. Data augmentation refers to the practice of synthetically creating additional data items by modifying or recombining existing items. In the audio domain, this could be done for example by adding noise, filtering, or mixing audio clips together \cite{Lasseck:2018birdclef}.
However, simple unprincipled data augmentation does not reduce issues such as undersampling (e.g. some vocalisations unrepresented in data set) or confounding factors.

There thus remains a gap in applying machine learning for automatic individual identification as a general-purpose tool that can be shown to be reliable for multiple species and can generalise correctly across recording conditions.

In the work reported in this paper, we tested generalisation of machine learning across species and across recording conditions in context of individual acoustic identification.
We used extensive data for three different bird species, including repeated recordings of the same individuals within and across two breeding seasons.
As well as directly evaluating across seasons, we also introduced ways to modify the evaluation data to probe the generalisation properties of the classifier.
We then improved on the baseline approach by developing novel methods which help to improve generalisation performance, again by modifying the data used.
Although tested with selected species and classifiers, our approach of modifying the data rather than the classification algorithm was designed to be compatible with a wide variety of automatic identification workflows.

\section{Materials and methods}

\subsection{Data collection}
\begin{figure*}[t]
	\centering
	\includegraphics[page=1,width=0.99\linewidth,clip,trim=0mm 140mm 0mm 60mm]{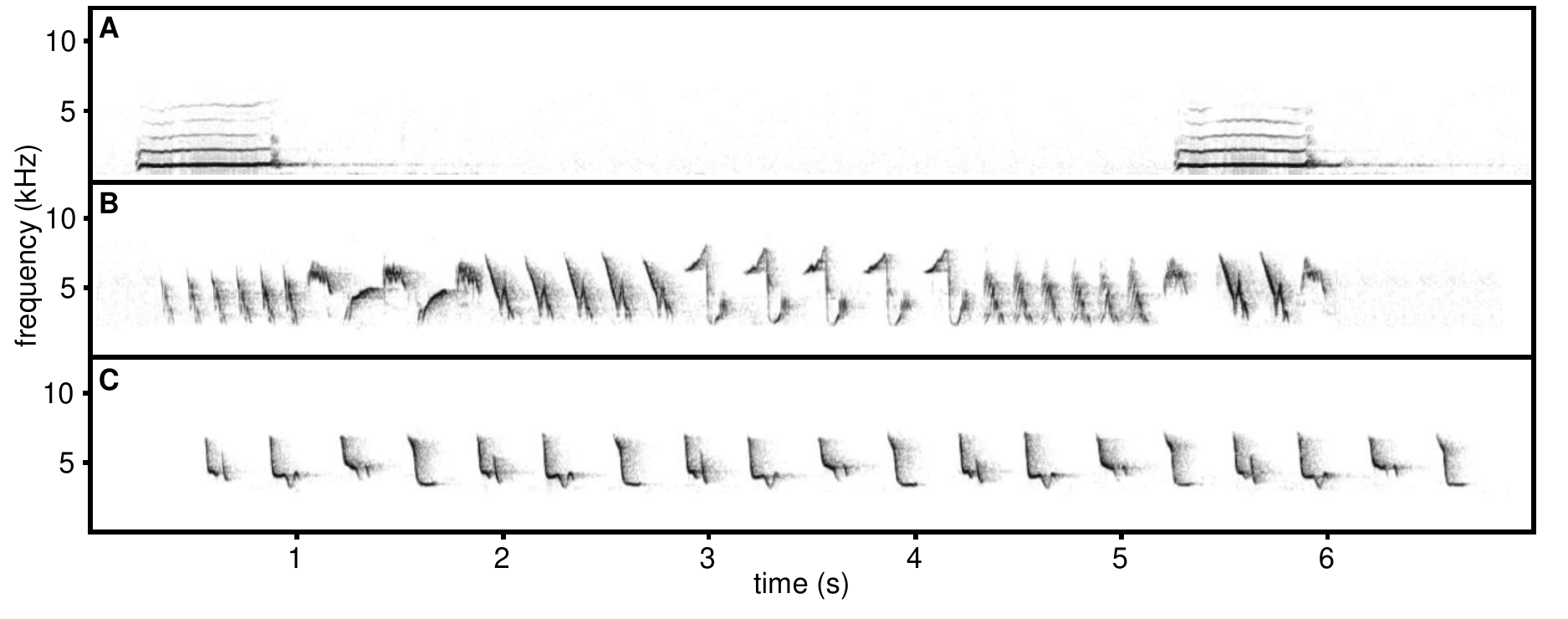}%
	\caption{Example spectrograms representing our three study species: \newline (a) little owl (b) tree pipit (c) chiffchaff.}
	\label{fig:spectrograms}
\end{figure*}

%
For this study we chose three bird species of varying vocal complexity (Figure \ref{fig:spectrograms}), in order to explore how a single method might apply to the same task at differing levels of difficulty and variation.
Little owl ({\textit{Athene noctua}}) represents a species with simple vocalisation (Figure \ref{fig:spectrograms}a): territorial call is a single syllable which is individually unique and it is held to be  stable over time (Linhart and \v{S}\'{a}lek\ unpubl. data) as was shown in several other owl species (e.g. \cite{delport_vocal_2002,grava_individual_2008}). Then, we selected two passerine species, which exhibit vocal learning: chiffchaff (\textit{Phylloscopus collybita}) and tree pipit (\textit{Anthus trivialis}). Tree pipit songs are also individually unique and stable over time \cite{Petruskova:2015}; but male on average uses 11 syllable types (6-18)  which are repeated in phrases that can be variably combined to create a song (\cite{Petruskova:2008}, Figure \ref{fig:spectrograms}b).
Chiffchaff song, when visualised, may seem simpler than that of the pipit. However, the syllable repertoire size might actually be higher---9 to 24 types---and, contrary to the other species considered, chiffchaff males may change syllable composition of their songs over time (\cite{Pruchova:2017}, (Figure \ref{fig:spectrograms}c). 
Selected species also differ in their ecology. While little owls are sedentary and extremely faithful to their territories \cite{nieuwenhuyse_little_2008}, tree pipits and chiffchaffs belong to migratory species with high fidelity to their localities. Annual returning rates for both are 25\%\ to  30\%\ (\cite{Petruskova:2015}, Linhart unpubl. data).

For each of these species, we used targeted recordings of single vocally active individuals. Distance to the recorded individual varied across individuals and species according to their tolerance towards people. We tried to get the best recording and minimise distance to each singing individual without disturbing its activities. Recordings were always done under favorable weather conditions (no rain, no strong wind). In general, signal-to-noise ratio is very good in all of our recordings (not rigorously assessed), but there are also environmental sounds, sounds from other animals or conspecifics in the recording background. All three species were recorded with following equipment: Sennheiser ME67 microphone, Marantz PMD660 or 661 solid-state recorder (sampling frequency 44.1 kHz, 16 bit, PCM). 

\textbf{Little owl (Linhart and \v{S}\'{a}lek 2017) \cite{Linhart:2017}:}
Little owls were recorded in two Central European farmland areas: northern Bohemia, Czech Republic (50{\textdegree}23’N, 13{\textdegree}40’E), and eastern Hungary (47{\textdegree}33ˈN, 20{\textdegree}54ˈE). Recordings were made from sunset until midnight between March and April of 2013---2014. Territorial calls of each male were recorded for up to three minutes after a short playback provocation (1 min) inside their territories from up to 50 m distance from the individuals. Identity of the males could not be explicitly checked because only a small proportion of males were ringed. Therefore, we inferred identity by the territory location combined with the call frequency modulation pattern which is distinctive per individual.

\textbf{Chiffchaff (Pr\r{u}chov\'{a} et al 2017 \cite{Pruchova:2017}, Pt\'{a}\v{c}ek et al 2016 \cite{Ptacek:2016}):}
Chiffchaff males were recorded in a former military training area on the outer boundary of \v{C}esk\'{e} Bud\v{e}jovice town, the Czech Republic (48{\textdegree}59.5ˈN, 14{\textdegree}26.5ˈE). Males were recorded for the purposes of various studies from 2008 up to and including 2011. Recordings were done from 05:30 to 11:00 hours in the morning. Only spontaneously singing males were recorded from within about 5--15 m distance. Identity of males was confirmed by colour rings.

\textbf{Tree Pipit (Petruskov\'{a} et al. 2015 \cite{Petruskova:2015}):}
Tree Pipit males were recorded at the locality Brdsk\'{a} vrchovina, the Czech Republic (49{\textdegree}84N, 14{\textdegree}10E) where the population has been continuously studied since 2011. Spontaneously singing males were recorded throughout whole day according to the natural singing activity of Tree pipits from mid-April to mid-July. Males were identified either based on colour ring observations or their song structure \cite{Petruskova:2015}.

Audio files were divided into regions during which the focal individual was vocally active (``foreground'') and inactive (``background'').
The total numbers of individuals and sound files in each dataset are summarised in Table \ref{tbl:datasets}.

\begin{table*}[t]
	\caption{Details of the audio recording datasets used.}
	\label{tbl:datasets}
\centering
\begin{tabular}{lccccc}
Evaluation scenario   &  \thead{Num.\ of \\ inds}  &
\thead{Foreground \\ \# audio files \\ (train : eval)} &  \thead{Foreground \\ total minutes \\ (train : eval)}  &
\thead{Background \\ \# audio files \\ (train : eval)} &  \thead{Background \\ total minutes \\ (train : eval)}  \\
\hline

Chiffchaff within-year         & 13 &   5107 : 1131  &  451 : 99  &  5011 : 1100  & 453 : 92  \\
Chiffchaff only-15             & 13 &  \ 195 : 1131  &   18 : 99  & \ 195 : 1100  &  21 : 92  \\
Chiffchaff across-year         & 10 &    324 : 201   &   32 : 20  &   304 : 197   &  31 : 24  \\
Little owl across-year         & 16 &    545 : 407   &   11 : 8   &   546 : 409   &  34 : 27  \\
Pipit within-year              & 10 &    409 : 303   &   27 : 21  &   398 : 293   &  49 : 47  \\
Pipit across-year              & 10 &    409 : 313   &   27 : 19  &   398 : 306   &  49 : 37  \\

\end{tabular}
\end{table*}

\subsection{Structured data augmentation}

``Data augmentation'' in machine learning refers to creating artificially large or diverse data sets by synthetically manipulating items in data sets to create new items---for example, by adding noise or performing mild distortions.
These artificially enriched data sets, used for training, often lead to improved automatic classification results, helping to mitigate the effects of limited data availability \cite{Krizhevsky:2012,Ciresan:2012}.
Data augmentation is increasingly used in machine learning applied to audio.
Audio-specific manipulations used might include filtering or pitch-shifting, or the mixing together of audio files (i.e.\ summing their signals together) \cite{Schluter:2015,Salamon:2017}.
Some of the highest-performing automatic species recognition systems rely in part on such data augmentations to attain their strongest results
\cite{Lasseck:2018birdclef}. 

\begin{figure*}[thp]
	\centering
\subfloat[`Foreground' recordings, which also contain some signal content coming from the background habitat. The foreground and background might not vary independently, especially in the case of territorial animals.]{
	\label{fig:sdapicsfg}
	\includegraphics[width=0.3\textwidth,clip,trim=0mm 0mm 0mm 0mm]{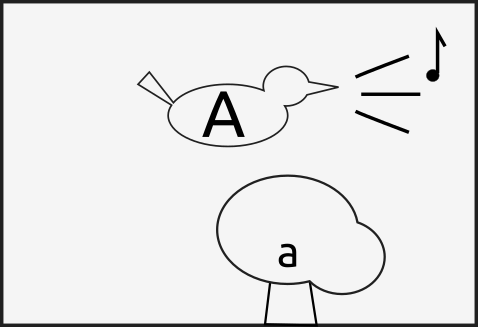}
	\qquad
	\includegraphics[width=0.3\textwidth,clip,trim=0mm 0mm 0mm 0mm]{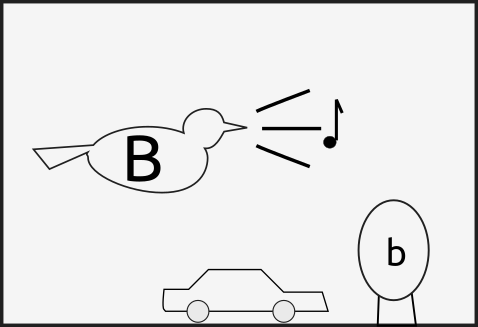}
	\hspace{2.0cm}
}

\subfloat[`Background' recordings, recorded when the focal animal is not vocalising]{
	\label{fig:sdapicsbg}
	\includegraphics[width=0.3\textwidth,clip,trim=0mm 0mm 0mm 0mm]{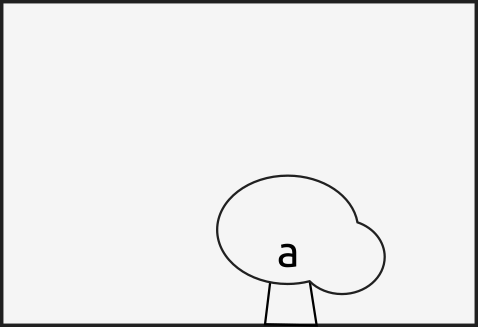}
	\qquad
	\includegraphics[width=0.3\textwidth,clip,trim=0mm 0mm 0mm 0mm]{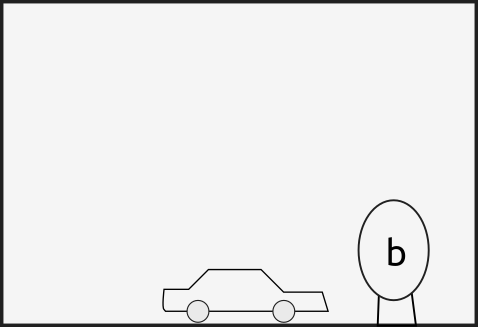}%
	\hspace{2.0cm}
}

\subfloat[In \textit{adversarial} data augmentation, we mix each foreground recording with a background recording from another individual, and measure the extent to which this alters the classifier's decision.]{
	\label{fig:sdapicsadv}
	\includegraphics[width=0.3\textwidth,clip,trim=0mm 0mm 0mm 0mm]{images/illustBb}%
	\raisebox{12mm}{$\rightarrow${classify}}%
	\qquad\raisebox{12mm}{vs.}\qquad
	\includegraphics[width=0.3\textwidth,clip,trim=0mm 0mm 0mm 0mm]{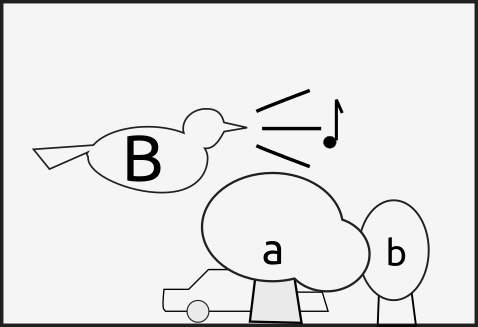}
	\raisebox{12mm}{$\rightarrow${classify}}%
}

\subfloat[In \textit{stratified} data augmentation, each foreground recording is mixed with a background recording
\textit{from each other class}. This creates a  to reduce the confounding correlation in the training data.]{
	\label{fig:sdapicsstrat}
	\includegraphics[width=0.3\textwidth,clip,trim=0mm 0mm 0mm 0mm]{images/illustBab}
	\quad
	\includegraphics[width=0.3\textwidth,clip,trim=0mm 0mm 0mm 0mm]{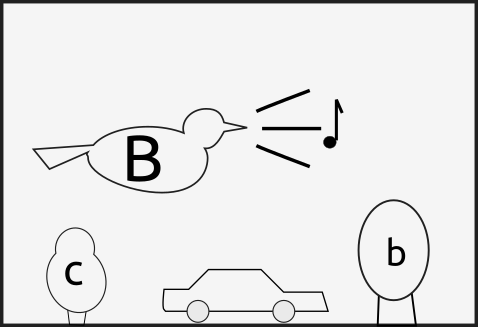}
	\quad\raisebox{12mm}{...\qquad$\rightarrow${train}}%
}

\caption{Explanatory illustration of our data augmentation interventions.}
\label{fig:sdapics}
\end{figure*}

In this work, we describe two augmentation methods used specifically to evaluate and to reduce the confounding effect of background sound. These \textit{structured} data augmentations are based on audio mixing but with the combinations of files to mix selected based on foreground and background identity metadata.
We make use of the fact that when recording audio from focal individuals in the wild, it is common to obtain recording clips in which the focal individual is vocalising (Figure \ref{fig:sdapicsfg}),
as well as `background' recordings in which the vocal individual is silent (Figure \ref{fig:sdapicsbg}).
The latter are commonly discarded. We used them as follows:

\begin{description}
\item[Adversarial data augmentation:]
To evaluate the extent to which confounding from background information is an issue,
we created datasets in which each foreground recording has been mixed with one background recording from some other individual
(Figure \ref{fig:sdapicsadv}).
In the best case, this should make no difference, since the resulting sound clip is acoustically equivalent to a recording of the foreground individual, but with a little extra irrelevant background noise.
In fact it could be considered a synthetic test of the case in which an individual is recorded having travelled out of their home range.
In the worst case, a classifier that has learnt undesirable correlations between foreground and background will be misled by the modification, either increasing the probability of classifying as the individual whose territory provided the extra background, or simply confusing the classifier and reducing its general ability to classify well.
In our implementation, each foreground item was used once, each mixed with a different background item. Thus the evaluation set remains the same size as the unmodified set.
We evaluated the robustness of a classifier by looking at any changes in the overall correctness of classification, or in more detail via the extent to which the classifier outputs are modified by the adversarial augmentation.

\item[Stratified data augmentation:]
We can use a similar principle during the training process, to create an enlarged and improved training data set.
We created training datasets in which each training item had been mixed with an example of background sound from each other individual (Figure \ref{fig:sdapicsstrat}).
If there are $K$ individuals this means that each item is converted into $K$ synthetic items, and the data set size increases by a factor of $K$.
Stratifying the mixing in this way, rather than selecting background samples purely at random, is intended to expose a classifier to training data with reduced correlation between foreground and background, and thus reduce the chance that it uses confounding information in making decisions.
\end{description}

To implement the foreground and background audio file mixing, we used the \texttt{sox} processing tool v14.4.1.%

\subsection{Using background items directly}

Alongside our data augmentation, we can also consider simple interventions in which the background sound recordings are used alone without modification.

One way of diagnosing confounding-factor issues in AAII is to apply the classifier to \textit{background-only} sound recordings. If there are no confounds in the trained classifier, trained on foreground sounds, then it should be \textit{unable} to identify the corresponding individual for any given background-only sound (identifying `a' or `b' in Figure \ref{fig:sdapicsbg}). Automatic identification (``AAII'') for background-only sounds should yield results at around chance level.

A second use of using the background-only recordings is to create an explicit `wastebasket' class during training.
As well as training the classifier to recognise individual labels A, B, C, ..., we created an additional `wastebasket' class which should be recognised as `none of the above', or in this case, explicitly as `background'.
The explicit-background class may or may not be used in the eventual deployment of the system.
Either way, its inclusion in the training process could help to ensure that the classifier learns not to make mistaken associations with the other classes.
This approach is related to the universal background model (UBM) used in open-set recognition methods \cite{Ptacek:2016}.
Note that the `background' class is likely to be different in kind from the other classes, having very diverse sounds.
In methods with an explicit UBM, the background class can be handled differently than the others \cite{Ptacek:2016}.
Here, we chose to use methods that can work with any classifier, and so the background class was simply treated analogously to the classes of interest.

\subsection{Automatic classification}

\begin{figure*}[thp]
\subfloat[A standard workflow for automatic audio classification. The upper portion shows the training procedure, and the lower shows the application or evaluation procedure.]{
\begin{centering}
	\includegraphics[page=1,width=0.49\linewidth,clip,trim=0mm 0mm 0mm 0mm]{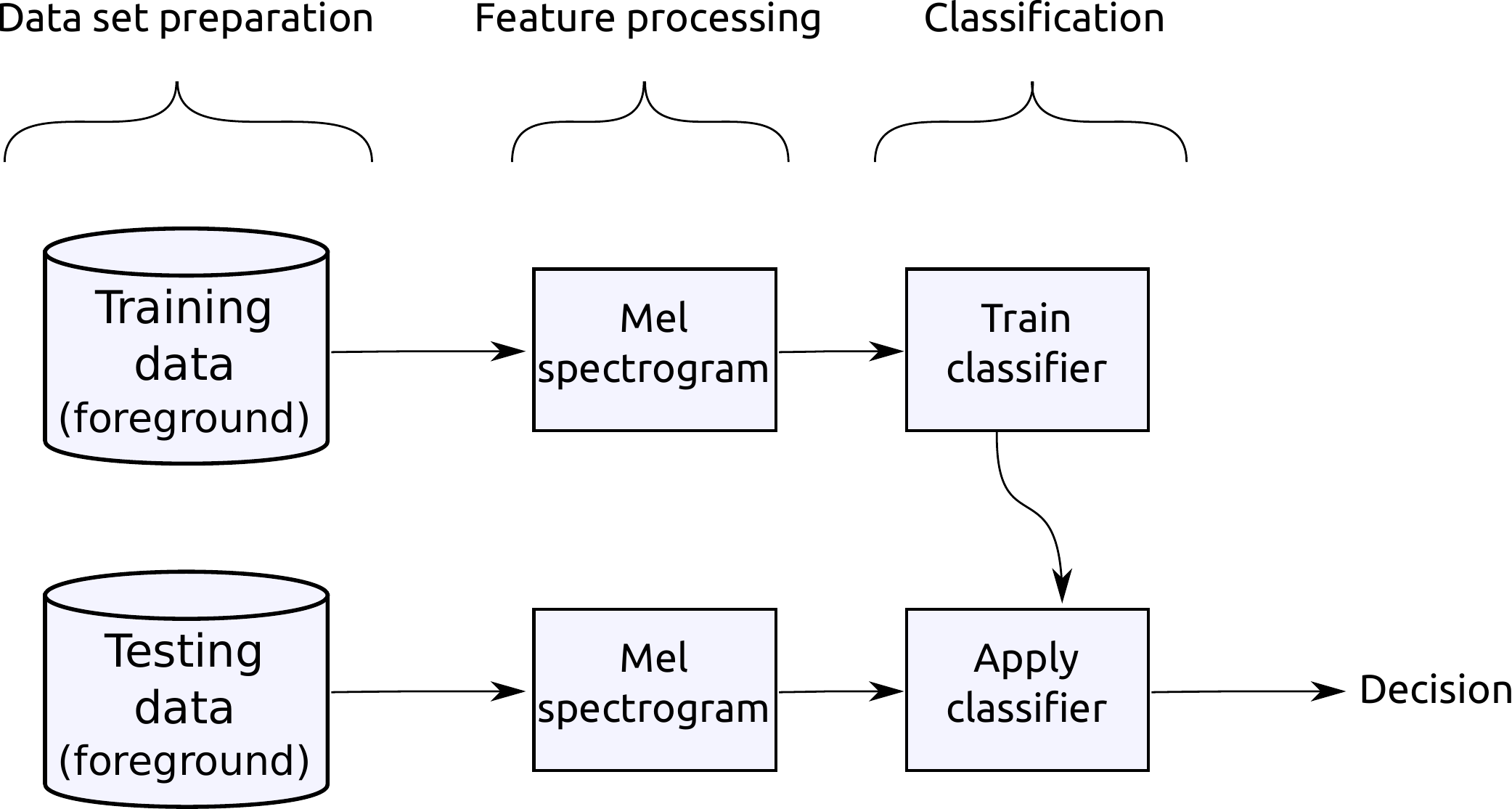}%
\end{centering}
	\label{fig:classifworkflowbasic}
}

\subfloat[Workflow for our automatic classification experiments. Dashed boxes represent steps which we enable/disable as part of our experiment. The upper portion shows the training procedure, and the lower shows the evaluation procedure. The two portions are very similar. However, note that the purpose and method of augmentation is different in each, as is the use of background-only audio: in the training phase the `concatenation' block creates an enlarged training set as the union of the background items and the foreground items, while in the evaluation phase the `choose' block select only one of the two, for the system to make predictions about.]{
	\centering
	\includegraphics[page=1,width=0.99\linewidth,clip,trim=0mm 0mm 0mm 0mm]{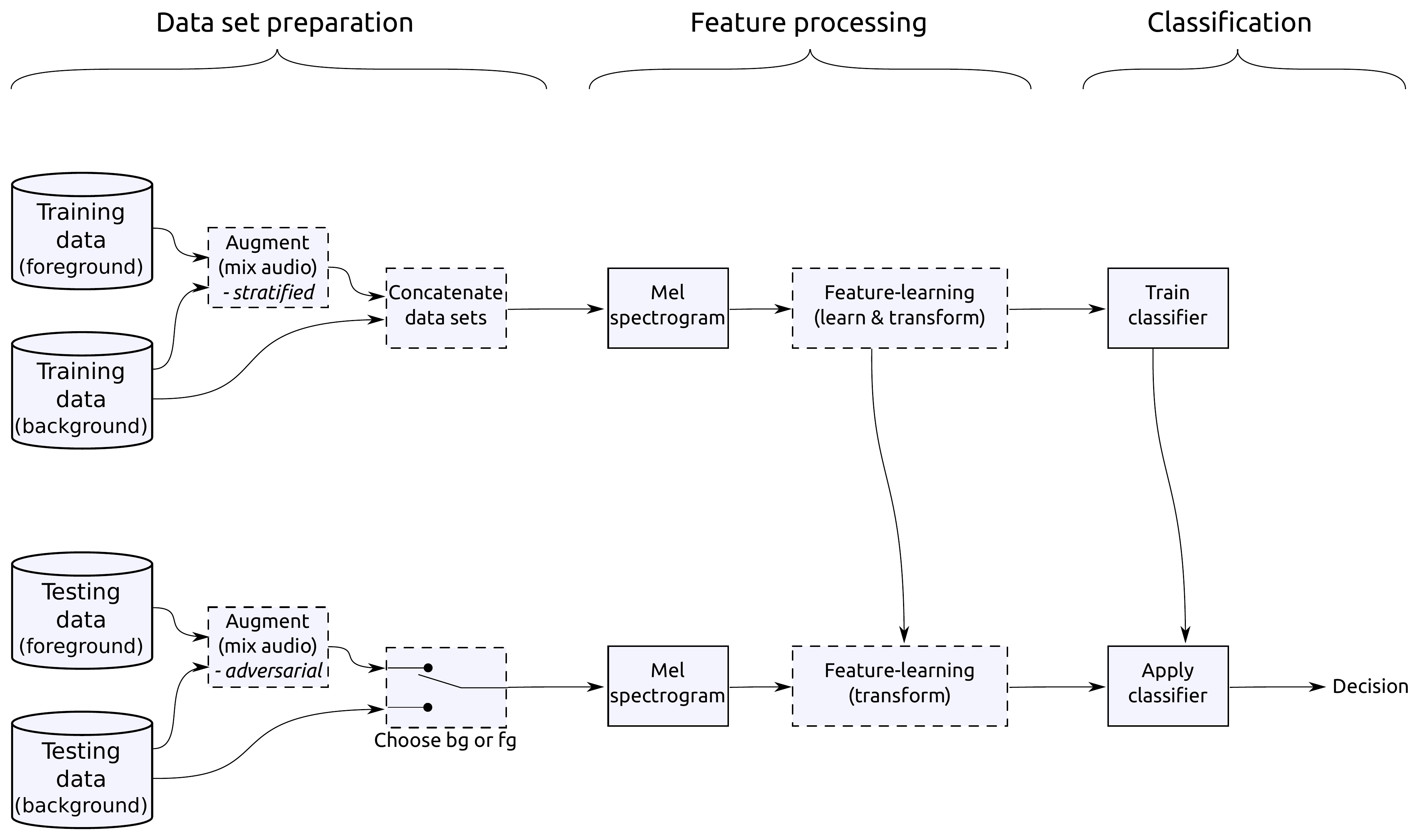}%
	\label{fig:classifworkflow}
}
\caption{Classification workflows.}
\end{figure*}

In this work, we started with a standard automatic classification processing workflow (Figure \ref{fig:classifworkflowbasic}),
and then experimented with inserting our proposed improvements. We modified the feature processing stage,
but our main innovations in fact came during the data set preparation stage, using the foreground and/or background data sets in various combinations to create different varieties of training and testing data (Figure \ref{fig:classifworkflow}).

As in many other works, the audio files---which in this case may be the originals or their augmented versions---were not analysed in their raw waveform format, but were converted to a mel spectrogram representation: `mel' referring to a perceptually-motivated compression of the frequency axis of a standard spectrogram.
We used audio files (44.1 kHz mono) converted into spectrograms using frames of length 1024 (23 ms), with Hamming windows, 50\% frame overlap, and 40 mel bands.
We applied median-filtering noise reduction to the spectrogram data.

Following the findings of \cite{Stowell:2014b}, we also applied \textit{unsupervised feature learning} to the mel spectrogram data as a preprocessing step.
This procedure scans through the training data in unsupervised fashion (i.e.\ neglecting the data labels), finding a linear projection that provides an informative transformation of the data.
We evaluated the audio feature data with and without this feature learning step,
to evaluate whether the data representation had an impact on the robustness and generalisability of automatic classification.
In other words, as input to the classifier we used either the mel spectrograms, or the learned representation obtained by transforming the mel spectrogram data.

The automatic classifier we used was one based on a random forest classifer that was previously tested successfully for bird species classification, but had not been tested for AAII \cite{Stowell:2014b}.

\subsection{Evaluation}

As is standard in automatic classification evaluation, we divided our datasets into portions used for training the system, and portions used for evaluating system performance.
Items used in training were not used in evaluation, and the allocation of items to the training or evaluation sets was done to create a partitioning through time: evaluation data came from different days within the breeding season, or subsequent years, than the training data.
This corresponds to a plausible use-case in which a system is trained with existing recordings and then deployed; the partitioning also helps to reduce the probability of over-estimating performance.

To quantify performance we used receiver operating curve (ROC) analysis, and as a summary statistic the area under the ROC curve (AUC).
The AUC summarises classifier performance and has various desirable properties for evaluating classification \cite{Fawcett:2006}.

We evaluated the classifiers following the standard paradigm used in machine learning.
Note that during evaluation, we optionally modified the evaluation data sets in two possible ways, as already described:
adversarial data augmentation, and background-only classification.
In all cases we used AUC as the primary evaluation measure.
However, we also wished to probe the effect of adversarial data augmentation in finer detail:
even when the overall decisions made by a classifier are not changed by modifying the input data,
there may be small changes in the full set of probabilities it outputs.
A classifier that is robust to adversarial augmentation should be one whose probabilities change little if at all.
Hence for the adversarial augmentation test, we also took the probabilities output from the classifier and compared them against their equivalent probabilities from the same classifier in the non-adversarial case.
We measured the difference between these sets of probabilities simply by their root-mean-square error (RMS error).

\subsection{Phase one: testing with chiffchaff}
For our first phase of testing, we wished to compare the effectiveness of the different proposed interventions, and their relative effectiveness on data tested within-year or across-year. We chose to use the chiffchaff datasets for these tests, since the chiffchaff song has an appropriate level of complexity to elucidate the differences between classifier performance, in particular the possible change of syllable composition across years. The chiffchaff dataset is also by far the largest.

We wanted to explore the difference in estimated performance when evaluating a system with recordings from the same year, separated by days from the training data, versus recordings from a subsequent year.
In the latter case, the background sounds may have changed intrinsically, or the individual may have moved to a different territory;
and of course the individual's own vocalisation patterns may change across years.
This latter effect may be an issue for AAII with a species such as the chiffchaff, and also impose limits to the application of previous approaches such as template-based matching.
Hence we wanted to test whether this more flexible machine learning approach could detect individual signature in the chiffchaff even when applied to data from a different field season.
We thus evaluated performance on `within-year' data---recordings from the same season---and `across-year' data---recordings from the subsequent year, or a later year.

Since the size of data available is often a practical constraint in AAII, and since dataset size can have a strong influence on classifier performance,
we further performed a version of the `within-year' test in which the training data had been restricted to only 15 items per individual.
The evaluation data was not restricted.

To evaluate formally the effect of the different interventions, we applied generalised linear mixed models (GLMM) to our evaluation statistics, using the \texttt{glmmadmb} package within R version 3.4.4 \cite{Fournier:2012,Skaug:2016}.
Since AUC is a continuous value constrained to the range $[0,1]$, we used a beta link function.
Since RMSE is a non-negative error measure, we used a gamma family with a logarithmic link function.
For each of these two evaluation measures, we applied a GLMM, using the data from all three evaluation scenarios (within-year, cross-year, only-15).
The evaluation scenario was included as a random effect.
Since the same evaluation-set items were reused in differing conditions, this was a repeated-measures model with respect to the individual song recordings.

\subsection{Phase two: testing multiple species}

In the second phase of our investigations, we evaluated the selected approach across the three species separately:
chiffchaff, pipit and little owl.
For each of these we compared the most basic version of the classifier (using mel features, no augmentation, and no explicit-background)
against the improved version that was selected from phase one of the investigation.
For each species separately, and using within-year and across-year data according to availability,
we evaluated the basic and the improved classifier for the overall performance (AUC measured on foreground sounds).
We also evaluated their performance on background-only sounds, and on the adversarial data augmentation test,
both of which checked the relationship between improved classification performance and improvements or degradations in the handling of confounding factors.

For both of these tests (background-only testing and adversarial augmentation),
we applied GLMM tests similar to those already stated. In these cases we entered separate factors for the testing condition and for whether the improved classifier was in use, as well as an interaction term between the two factors.
This therefore tested for an effect of whether our improved classifier indeed mitigated the problems that the tests were designed to expose.

\section{Results}

\subsection{Phase one: chiffchaff}

\begin{figure*}[t]
	\centering
	\includegraphics[page=1,width=0.69\linewidth,clip,trim=0mm 0mm 0mm 0mm]{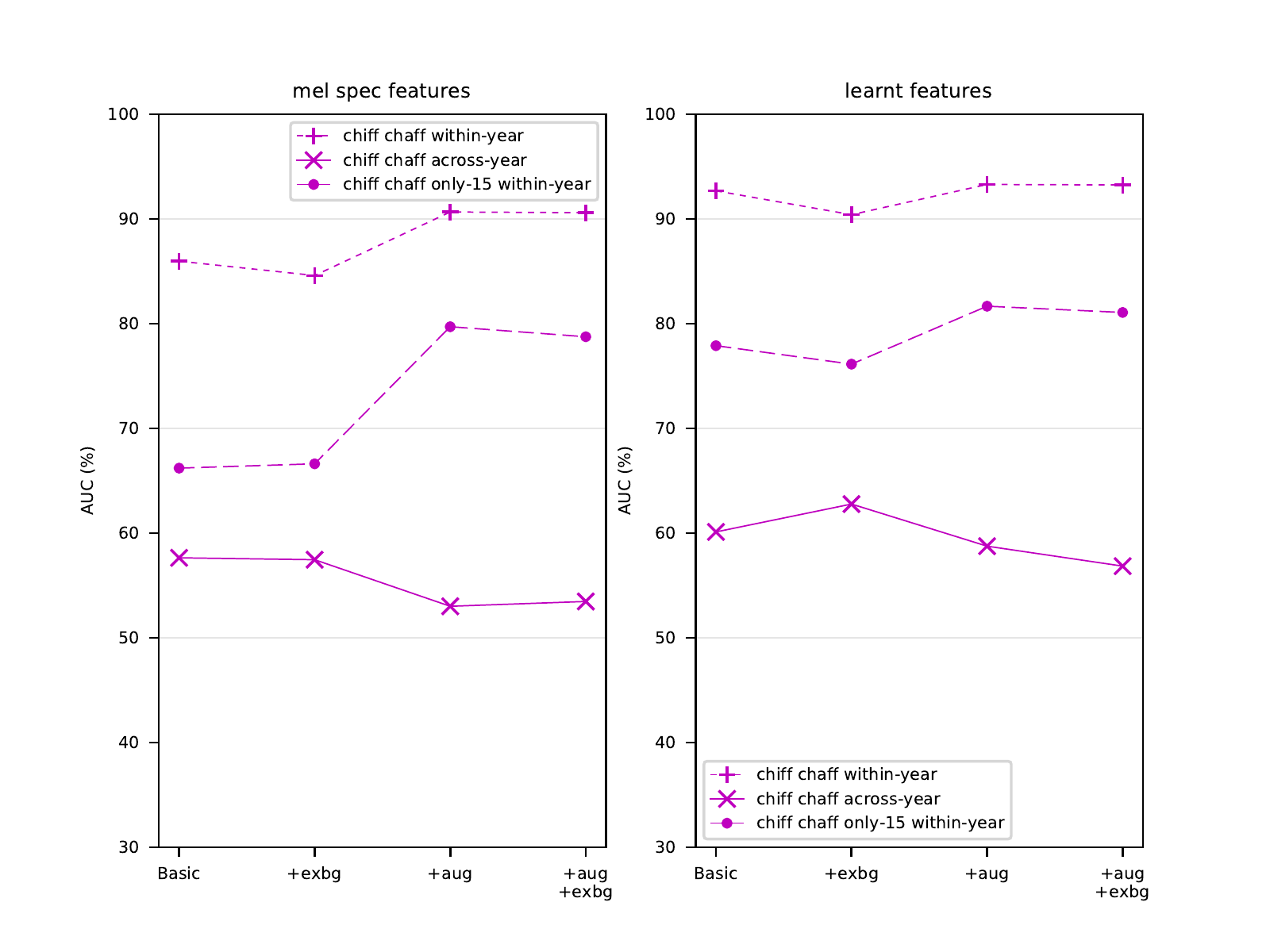}%
	\caption{Performance of classifier (AUC) across the three chiffchaff evaluation scenarios, and with various combinations of configuration: with/without augmentation (`aug'), learnt features, and explicit-background (`exbg') training.}
%
%
	\label{fig:chchauc}
\end{figure*}

AAII performance over the 13 chiffchaff individuals was strong, above 85\% AUC in all variants of the within-year scenario (Figure \ref{fig:chchauc}).
For interpretation, note that this corresponds to over 85\% probability that a random true-positive item is ranked higher than a random true-negative item by the system \cite{Fawcett:2006}.
This reduced to around 70--80\% when the training set was limited to 15 items per individual,
and reduced even further to around 60\% in the across-year evaluation scenario.
Recognising chiffchaff individuals across years remains a challenging task even under the studied interventions.

The focus of our study is on discriminating between individuals, but our ``explicit-background'' configuration additionally made it possible for the same classifier to discriminate between cases where a focal individual was singing, and cases where it was not. Across all three of the conditions mentioned above, foreground-vs-background discrimination (aka ``detection'' of any focal individual) for chiffchaff was strong at over 95\% AUC. Mel spectral features performed slightly better for this (range 96.6--98.6\%) than learnt features (range 95.3--96.7\%). Given this, in the remainder of the results we focus on our main question of discriminating between individuals.

We tested the GLMM residuals for the two evaluation measures (AUC, RMSE) and found no evidence for overdispersion.
We also tested all possible reduced models with factors removed, comparing among models using AIC.
In both cases, the full model as well as a model with `exbg' (explicit-background training) removed gave the best fit,
with the full model less than 2 units above the exbg-reduced model and leading to no difference in significance estimates.
We therefore report results from the full models.

Feature-learning and structured data augmentation were both found to significantly improve
classifier performance (Table \ref{tbl:glmmauc})
as well as robustness to adversarial data augmentation (Table \ref{tbl:glmmrmse}).
Explicit-background training was found to lead to mild improvement but this was a long way below significance.

\begin{table*}[t]
	\caption{Results of GLMM test for AUC, across the three chiffchaff evaluation scenarios.}
	\label{tbl:glmmauc}
\centering
\begin{tabular}{lll}
                          & Estimate & p-value   \\
\hline
(Intercept)               & 0.8199    & 0.041 *  \\
Feature-learning          & 0.3093    & 0.014 *  \\
Augmentation              & 0.2509    & 0.048 *  \\
Explicit-bg class         & 0.0626    & 0.621    \\
\end{tabular}
\end{table*}

\begin{table*}[th]
	\caption{Results of GLMM fit for RMSE in the adversarial data augmentation test, across the three chiffchaff evaluation scenarios.}
	\label{tbl:glmmrmse}
\centering
\begin{tabular}{lll}
                          & Estimate & p-value   \\
\hline
(Intercept)               &  1.8543    & 1.9e-05 *** \\
Feature-learning          & -0.5044    & 1.9e-08 *** \\
Augmentation              & -0.8734    & $<$ 2e-16 *** \\
Explicit-bg class         & -0.0141    & 0.87     \\
\end{tabular}
\end{table*}

\subsection{Phase two: multiple species}

Based on the results of our first study, we took forward an improved version of the classifier (using stratified data augmentation, and learnt features, but not explicit-background training) to test across multiple species.

\begin{figure*}[th]
	\centering
	\includegraphics[page=3,width=0.49\linewidth,clip,trim=0mm 0mm 85mm 0mm]{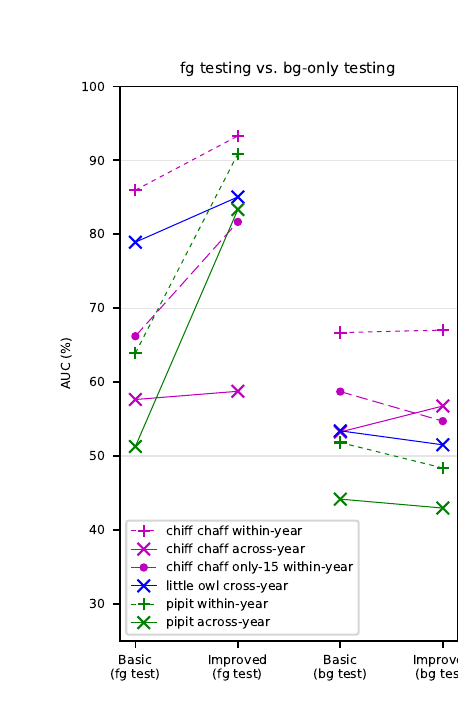}%
	\caption{Our selected interventions---data augmentation and feature-learning---improve classification performance, in some cases dramatically (left-hand pairs of points), without any concomitant increase in the background-only classification (right-hand pairs of points) which would be an indication of counfounding.}
	\label{fig:bgonlyauc}
\end{figure*}

Applying this classifier to the different species and conditions, we found that it led in most cases to a dramatic improvement in recognition performance of foreground recordings, and little change in the recognition of background recordings (Figure \ref{fig:bgonlyauc}, Table \ref{tbl:glmmaucfgbg}).
This suggests that the improvement is based on the individuals' signal characteristics and not confounding factors.

\begin{figure*}[th]
	\centering
	\includegraphics[page=4,width=0.49\linewidth,clip,trim=0mm 0mm 85mm 0mm]{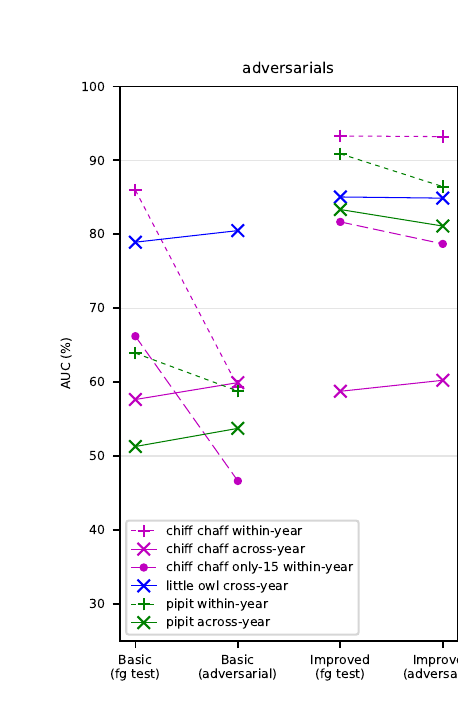}%
	\caption{Adversarial augmentation has a varied impact on classifier performance (left-hand pairs of points), in some cases giving a large decline. Our selected interventions vastly reduce the impact of this adversarial test, while also generally improving classification performance (right-hand pairs of points).}
	\label{fig:adversarialauc}
\end{figure*}

Our adversarial augmentation, intended as a diagnostic test to adversarially reduce classification performance, did not have strong overall effects on the headline performance indicated by the AUC scores (Figure \ref{fig:adversarialauc}, Table \ref{tbl:glmmaucfgbg}).
Half of the cases examined---the across-year cases---were not adversely impacted, in fact showing a very small increase in AUC score.
The chiffchaff within-year tests were the only to show a strong negative impact of adversarial augmentation, and this negative impact was removed by our improved classification method.

\begin{figure*}[th]
	\centering
	\includegraphics[page=5,width=0.49\linewidth,clip,trim=0mm 0mm 0mm 0mm]{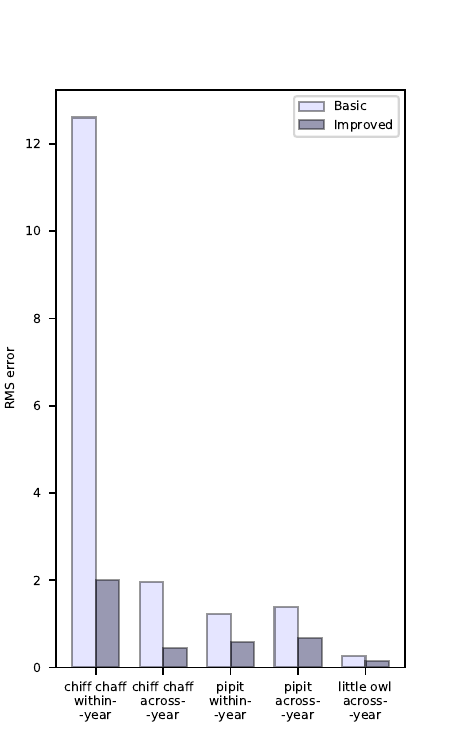}%
	\caption{Measuring in detail how much effect the adversarial augmentation has on classifier decisions: RMS error of classifier output, in each case applying adversarial augmentation and then measuring the differences compared against the non-adversarial equivalent applied to the exact same data. In all five scenarios, our selected interventions lead to a large decrease in the RMS error.}
%
%
	\label{fig:adversarialrmse}
\end{figure*}

We also conducted a more fine-grained analysis of the effect of augmentation, by measuring the amount of deviation induced in the probabilities output from the classifier. On this measure we observed a consistent effect, with our improvements reducing the RMS error by ratios of approx 2--6, while the overall magnitude of the error differed across species (Figure \ref{fig:adversarialrmse}).
%
%

\begin{table*}[th]
	\caption{Results of GLMM test for AUC, across all three species, to quantify the general effect of our improvements on the foreground test and the background test (cf.\ Figure \ref{fig:bgonlyauc}).}
	\label{tbl:glmmaucfgbg}
\centering
\begin{tabular}{lll}
                          & Estimate & p-value   \\
\hline
(Intercept)                 &  0.792    &  0.00150 **   \\
Use of improved classifier  &  0.852    &  0.00032 ***  \\
Background-only testing     & -0.562    &  0.00624 **   \\
Interaction term            & -0.896    &  0.00391 **   \\
\end{tabular}
\end{table*}

\begin{table*}[th]
	\caption{Results of GLMM test for AUC, across all three species, to quantify the general effect of our improvements on the adversarial test (cf.\ Figure \ref{fig:adversarialauc}).}
	\label{tbl:glmmaucadv}
\centering
\begin{tabular}{lll}
                          & Estimate & p-value   \\
\hline
(Intercept)                     &  0.873 &   0.0121 *    \\
Use of improved classifier      &  0.820 & 0.0027 **   \\
Adversarial data augmentation   & -0.333 & 0.1713      \\
Interaction term                &  0.225 & 0.5520      \\
\end{tabular}
\end{table*}

\clearpage
\section{Discussion}


We demonstrate that a single approach to automatic acoustic identification of individuals (AAII) can be successfully used across different species with different complexity of vocalisations. One exception to this is the hardest case, chiffchaff tested across years, in which automatic classification performance remains modest. The chiffchaff case (complex song, variable song content), in particular, highlights the need for proper assessment of identification performance. Without proper assessment we cannot be sure if promising results reflect the real potential of proposed identification method. We document that our proposed improvements to the classifier training process are able, in some cases, to improve the generalisation performance dramatically and, on the other hand, reveal confounds causing over--optimistic results.

We evaluated \textit{spherical k-means} feature-learning as previously used for species classification \cite{Stowell:2014b}.
We found that for individual identification it provides an improvement over plain Mel spectral features,
not just in accuracy (as previously reported) but also in resistance to
confounding factors (ibid.). We believe this is due to the feature-learning having
been tailored to reflect fine temporal details of bird sound; if so, this
lesson would carry across to related systems such as convolutional neural
networks. Our machine-learning approach may be particularly useful for automatic identification of individuals in species with more complex songs, such as pipits (note huge increase in performance over mel features in Figure \ref{fig:bgonlyauc}), or chiffchaffs (on short-time scale though).

Using silence-regions from focal individuals to create an ``explicit-background'' training category provided only a mild improvement in the behaviour of the classifier, under various evaluations. 
Also, we found that the best-performing configuration used for detecting the presence/absence of a focal individual was not the same as the best-performing configuration for discriminating between individuals.
Hence, it seems generally preferable not to combine the detection and AAII tasks into one classifier.

By contrast, using silence-regions to perform dataset augmentation of the foreground sounds was found to give a strong boost to performance as well as resistance against confounding factors. Background sounds are useful in training a system for AAII, through data augmentation (rather than explicit-background training).


We found that adversarial augmentation provided a useful tool to diagnose concerns about the robustness of an AAII system. In the present work we found that the classifier was robust against this augmentation (and thus we can infer that it was largely not using background confounds to make its decision), except for the case of chiffchaff with the simple mel features (Figure \ref{fig:adversarialauc}). This latter case exhorts us to be cautious, and suggests that results from previous call-type independent methods may have been over-optimistic in assessing performance \cite{Fox:2008,Fox:2008a,Cheng:2010,Cheng:2012,Ptacek:2016}.
Our adversarial augmentation method can help to test for this even in the absense of across-year data.

Background-only testing was useful to confirm that when the performance of a classifier was improved, the confounding factors were not aggravated in parallel, i.e. that the improvement was due to signal and not confound (Figure \ref{fig:bgonlyauc}). However, the performance on background sound recordings was not reduced to chance, but remained at some level reflecting the foreground-background correlations in each case, so results need to interpreted comparatively against the foreground improvement, rather than in isolation.
This individual specificity of the background may be related to the time interval between recordings.
This is clear from the across-year outcomes; within-year, we note that there was one day of temporal separation for chiffchaffs (close to 70 percent AUC on background-only sound), while an interval of weeks for pipits (chance-level classification of background).
These effects surely depend on characteristics of the habitat.

Our improved classifier performs much more reliably than the standard one; however, the most crucial factor still seems to be a targeted species.
For the little owl we found good performance, and least affected by modifications in methods - consistent with the fact that it is the species with the simplest vocalisations. Little owl represents a species well suited for template matching individual identification methods which have been used in past for many species with similar simple, fixed vocalisations (discriminant analysis, cross-correlation). For these cases, it seems that our automatic identification method does not bring advantage regarding improved classification performance. However, a general classifier such as ours, automatically adjusting a set of features for each species, would allow common users to start individual identification right away without the need to choose an appropriate template-matching method (e.g. \cite{Linhart:2017}). 

We found that feature learning gave the best improvement in case of pipits (Figure \ref{fig:bgonlyauc}). Pipits have more complex song, where simple template matching cannot be used to identify individuals. In pipits, each song may have different duration and may be composed of different subsets of syllable repertoire, and so any a single song cannot be used as template for template matching approach. This singing variation likely also prevents good identification performance based on Mel features in pipits. Nevertheless, a singing pipit male will cycle through the whole syllable repertoire within a relatively low number of songs and individual males can be identified based on their unique syllable repertoires (\cite{Petruskova:2015}). We think that our improvements to the automatic identification might allow the system to pick up correct features associated with stable repertoire of each male. This extends the use of the same automatic identification method to the large part of songbird species that organise songs into several song types and, at the same time, are so-called closed-ended learners (\cite{beecher_functional_2005}).      

Our automatic identification, however, cannot be considered fully independent of song content in a sense defined earlier (e.g.\cite{Fox:2008,Cheng:2010}). Such content-independent identification method should be able to classify across-year recordings of chiffchaffs in which syllable repertoires of males differ almost completely between the two years \cite{Pruchova:2017}. Due to vulnerability of Mel feature classification to confounds reported here and because performance of content independent identification has been only tested on short-term recordings, we believe that the concept of fully content-independent individual identification needs to be reliably demonstrated yet. 

Our approach seems to be definitely suitable for species with individual vocalisation stable over time, even if that vocalisation is complex---a very wide range of species---in general outdoor conditions. For such species it might be successfully used for individual automatic acoustic monitoring, although this needs to be tested at larger scale: in various species and in large populations. In future work these approaches should also be tested with `open-set' classifiers allowing for the possibility that new unknown individuals might appear in data.
This is well-developed in the ``universal background model'' (UBM) developed in GMM-based speaker recognition \cite{Ptacek:2016}, and future work in machine learning is needed to develop this for the case of more powerful classifiers. 

Important for further work in this topic is open sharing  of data in standard formats.
Only this way can diverse datasets from individuals be used to develop/evaluate automatic recognition that works across many taxa and recording conditions.

We conclude by listing the recommendations that emerge from this work for users of automatic classifiers, in particular for acoustic recognition of individuals:

\begin{enumerate}
\item
Record `background' segments, for each individual (class), and publish background audio samples alongside the trimmed individual audio samples. Standard data repositories can be used for these purposes (e.g.\ Dryad, Zenodo).

\item
Improve robustness by: \newline
     (a) suitable choice of input features; \newline
     (b) structured data augmentation, using background sound recordings.

\item
Probe your classifier for robustness by: \newline
     (a) background-only recognition: higher-than-chance recognition strongly implies confound; \newline
     (b) adversarial distraction with background: a large change in classifier outputs implies confound; \newline
     (c) across-year testing (if such data are available): a stronger test than within-year.

\item
Be aware of how species characteristics will affect recognition. The vocalisation characteristics of the species will influence the ease with which automatic classifiers can identify individuals. Songbirds whose song changes within and between seasons will always be harder to identify reliably - as is also the case in manual identification.

\item
Best practice is to test manual features and learned features since the generalisation and performance characteristics are rather different. In the present work we compare basic features against learned features; for a different example see \cite{Mouterde:2014}. Manual features are usually of lower accuracy, but with learned features more care must be taken with respect to confounds and generalisation.
\end{enumerate}

\newpage
\section*{Ethics}

Our study primarily involved only non-invasive recording of vocalising individuals. In the case of ringed individuals (all chiffchaffs and some tree pipits and little owls), ringing was done by experienced ringers (PL, M\v{S}, TP) who all held ringing licences at the time of study. Tree pipits and chiffchaff males were recorded during spontaneous singing. Only for little owls short playback recording (1 min) was used to provoke calling. Playback provocations as well as handling during ringing were kept as short as possible and we are not aware of any consequences for subjects' breeding or welfare.

\section*{Data Accessibility}

Our audio data and the associated metadata files are available online under the Creative Commons Attribution licence (CC BY 4.0) at 
\url{http://doi.org/10.5281/zenodo.1413495}

\section*{Competing Interests}

We have no competing interests.

\section*{Authors' Contributions}
DS and PL conceived and designed the study.
PL, TP and M\v{S} recorded audio.
PL processed the audio recordings into data sets.
DS carried out the classification experiments and performed data analysis.
DS, PL and TP wrote the manuscript.
All authors gave final approval for publication.

\section*{Funding}

DS was supported by EPSRC Early Career research fellowship EP/L020505/1.
PL was supported by the National Science Centre, Poland, under Polonez fellowship reg. no UMO-2015/19/P/NZ8/02507 funded by the European Union’s Horizon 2020 research and innovation programme under the Marie Skłodowska-Curie grant agreement No 665778.
TP was supported by the Czech Science Foundation (project P505/11/P572).M\v{S} was supported by the research aim of the Czech Academy of Sciences (RVO 68081766).


\newpage
\bibliographystyle{vancouver}
\bibliography{references}


\end{document}